\documentclass[twoside,english]{iopart}
\usepackage[T1]{fontenc}
\usepackage{geometry}
\usepackage{amsthm}
\usepackage{graphicx}

\makeatletter
\usepackage{iopams}
\usepackage{setstack}

\newcommand{\eqref}[1]{(\ref{#1})}

\makeatother

\usepackage{babel}
\begin{document}
\title{Exact densities of loops in O(1) dense loop model and of clusters
in critical percolation on a cylinder. }
\author{{\Large{}A.M. Povolotsky}}
\address{{\large{}Bogoliubov Laboratory of Theoretical Physics, Joint Institute
for Nuclear Research, 141980, Dubna, Russia}\\
{\large{}National Research University Higher School of Economics,
20 Myasnitskaya, 101000, Moscow, Russia}}
\ead{alexander.povolotsky@gmail.com}
\begin{abstract}
We obtain exact densities of contractible and non-contractible loops
in the O(1) model on a strip of the square lattice rolled into an
infinite cylinder of finite even circumference $L$. They are also
equal to the densities of critical percolation clusters  on  forty five degree rotated square lattice rolled into a cylinder, which do not
or do wrap around the cylinder respectively. The results are presented as explicit rational functions
of $L$ taking rational values for any even $L$. Their asymptotic
expansions in the large $L$ limit have irrational coefficients reproducing
the earlier results in the leading orders. The solution is based on
a mapping to the six-vertex model and the use of technique of Baxter's
T-Q equation.
\end{abstract}

\maketitle
\maketitle

\section{Introduction}

The subject of this Letter, $O(1)$ dense loop model (DLM), is a particular
case of $O(n)$ loop models, a class of lattice models of statistical
physics formulated in terms of ensembles of paths on the lattices.
Having connections with many other models they sometimes provide an
alternative convenient language for the analysis. An idea of representing
the partition function of the Ising model as a sum over sets of weighted
contours comes back to Peierls \cite{Peierls1936}. The summation over
contours with  weight $n$ assigned to loops appeared from the polygonal
representation of the partition function of the random cluster model
\cite{FK1972,BaxterKellandWu}, which, in turn, is related to the
$Q$-state Potts model with $\sqrt{Q}=n.$ Also, a connection of the
$O(n)$ loop model with the $O(n)$ vector model, from which the former
inherited the $O(n)$ name, suggests that the former can be used to
predict the critical behavior of the the latter \cite{DMNS}. The
language of $O(n)$ loop models turned out especially efficient within
the framework of Coulomb gas and conformal field theory (CFT) \cite{Nienhuis1987,Jacobsen},
while their scaling limit fits naturally into the Schramm-Loewner
evolution picture \cite{NienhuisKager2009}.

\begin{figure}[h]
\centering{}\includegraphics[width=0.3\textwidth]{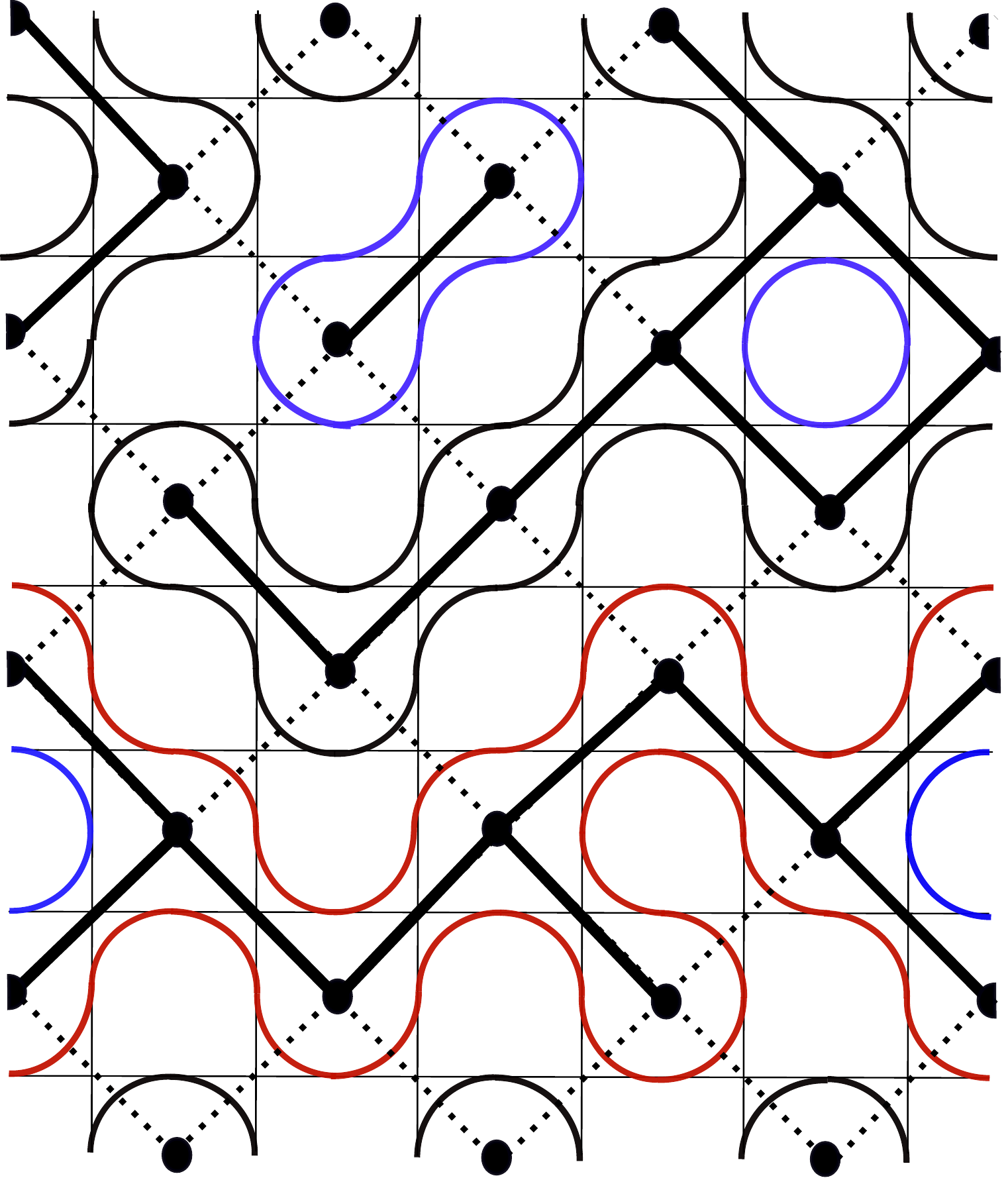}\caption{\label{fig:lattice}A configuration of the O(1) loop model on the
lattice (thin solid lines) and the associated percolation clusters
(thick solid lines) on the rotated lattice (dashed lines and black
dots). The contractible and non-contractible loops are shown in blue
and red respectively. }
\end{figure}

Here we consider the $O(1)$ DLM formulated as a measure on paths
on the two-dimensional square lattice. A path passes through every
bond exactly once, and two paths meet at every site without crossing
each other, see fig. \ref{fig:lattice}. All path configurations have
equal weights. 

\begin{figure}[h]
\centering{}\includegraphics[width=0.2\textwidth]{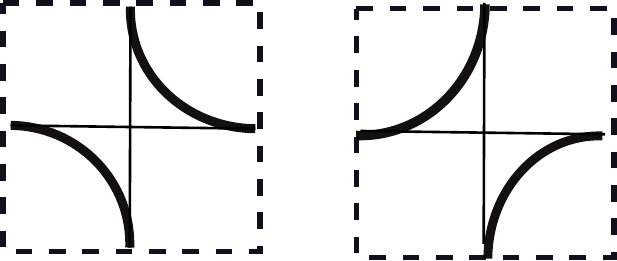}\caption{Two vertices of the O(1) loop models. Both vertices have unit weight.\label{fig:loop-vertices}}
\end{figure}

To construct configurations by local operations we place a vertex
at every lattice site, in which two pairs of paths at four incident
bonds are connected pairwise in one of two possible ways shown in
fig. \ref{fig:loop-vertices}. Both vertices  are assigned the unit
weight. 

The lattice we consider here is bounded in one spacial direction and
unbounded in the other. Specifically, it is a strip of the square
lattice, infinite in the direction that coincides with one of the
lattice directions, referred to as vertical, and finite in the other
(horizontal) direction with even number $L=2N$ of sites. Periodic
boundary conditions are implied in the horizontal direction, i.e.
the strip is rolled into a cylinder. Under the uniform measure on
paths only finite closed loops present on such a cylinder with probability
one, each loop having the weight $n=1$.

\begin{figure}
\centering{}\includegraphics[width=0.4\textwidth]{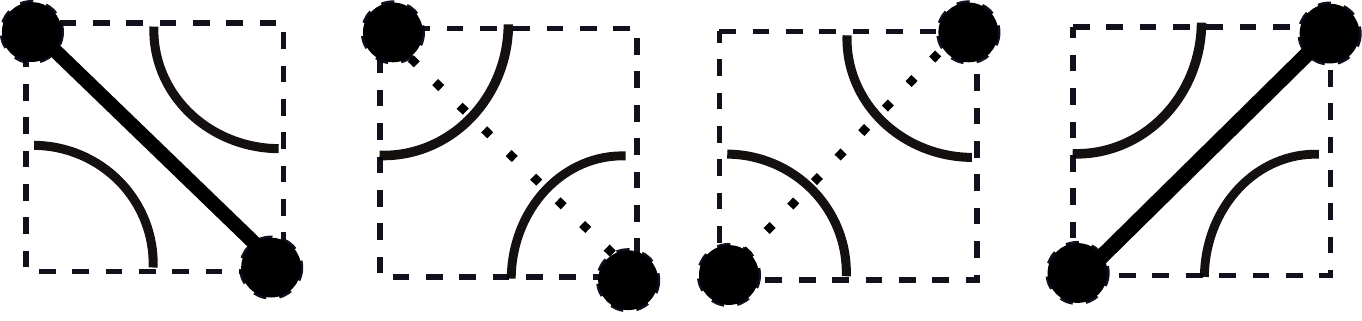}\caption{Correspondence between the vertices of O(1) DLM and open (solid) or
closed (dashed) bonds on the rotated lattice. The black dots are sites
of the rotated lattice. \label{fig:loop-perc}}
\end{figure}

This model has been intensively studied on its own and especially
in view of its connection with the critical bond percolation problem,
which is a particular instance of the random cluster model related
to a formal $Q\to1$ limit of the Potts model. To go from loops to
percolation we construct a new square lattice of 45 degrees rotated
orientation, for which the original lattice is the so called medial
graph. To this end, we put sites of the new lattice to the center
of every second face of the original lattice in a staggered way, connecting
them by bonds passing through the nearest sites of the original lattice
as shown in fig. \ref{fig:lattice}. The periodic boundary conditions
for the original lattice suggest that the rotated lattice is also
rolled into the cylinder. Then, we consider the bond percolation on
the lattice constructed. Specifically a bond of the rotated lattice
is said to be open, if it is between the loop arcs and closed if it
crosses them, see fig. \ref{fig:loop-perc}. All the four vertices
have equal weights, i.e. open and closed bonds have equal probabilities
$p=1/2$, which is the critical point of the  bond percolation on the
infinite square lattice. 

The studies of percolation having been continuously conducted since
the late fifties of the last century culminated in plenty brilliant
results, see \cite{Grimmett,Bollobas} and references therein. In
particular, the connection of percolation with $O(1)$ DLM, in turn
related to the exactly solvable six-vertex model, was found especially
useful for calculating some observables and critical exponents \cite{Baxter}.
In the context of this Letter we mention the calculation of the
density of critical percolation clusters on the infinite plane lattice,
i.e. in the $L\to\infty$ limit of our cylinder, performed in a seminal
paper \cite{TL}. The result was obtained in the form of an integral,
of which the approximate numerical value was provided. The exact value
of the integral was later presented in \cite{ZiffFinchAdamchik}
together with numerical evidences of universality of finite size corrections
to this quantity occurring in confined geometries. An explicit form
of the finite size corrections to the density of critical percolation
clusters on the strip and on the cylinder were conjectured in \cite{KlebanZiff}
using arguments based on the Coulomb gas technique and CFT. Specifically,
the corrections come from the conformal anomaly, the value of which
was found from the mapping of the $O(1)$ DLM to the Coulomb gas \cite{Nienhuis1987}. 

In the $O(1)$ DLM language the quantity related to the density of
percolation clusters is the density of loops. In fact, the latter
can be considered as a combination of two quantities, which can be
studied separately. Indeed, there are two types of loops on the cylinder,
contractible and non-contractible, which do not and do wind around
the cylinder respectively. Hence, we will be interested in the average
numbers of loops of both types per site of the lattice. We will use
the notations $\nu_{c}(L)$ and $\nu_{nc}(L)$ for the densities of
contractible and non-contractible loops respectively.

To explain the relation of loop densities to the densities of percolation
clusters, we note that every contractible loop is either circumscribed
on a percolation cluster that does not wrap around the cylinder or
is inscribed into a circuit inside a percolation cluster. The latter
loop can also be thought as circumscribed on the dual percolation
cluster on the dual rotated lattice. The critical point is self-dual.
This means that the average numbers of percolation clusters and of
dual percolation clusters are equal, and so are the average numbers
of the circumscribed and the inscribed loops. Thus, the average number
of non-contractible loops per unit length of the cylinder is twice
the average number of the critical percolation clusters on the rotated
lattice. Since the rotated lattice contains twice less sites per unit
length of the cylinder than the original lattice, the density of percolation
clusters not wrapping around the cylinder coincides with $\nu_{c}(L)$. 

Also every percolation cluster that wraps around the cylinder is bounded
by a pair of non-contractible loops and every non-contractible loop
runs along the boundary of such a cluster. Thus, similarly to the
above, we argue that $\nu_{nc}(L)$ coincides with the average number
of critical percolation clusters wrapping around the cylinder per
site of the rotated lattice. 

How the mean density of non-contractible loops in $O(n)$
model on a cylinder depends on loop weights was studied in \cite{AlcarazBrankovPriezzhev RittenbergRogozhnikov}.
There, the conformal anomaly being a function of the weights of contractible
and non-contractible loops was obtained from the finite size correction
to the energy of the XXZ chain with twisted boundary conditions \cite{ABB,HQB,DdV}
divided by a coefficient termed the sound velocity \cite{GehlenRittenberg}.
This allowed a determination of the finite size corrections to the
loop densities. 

Unfortunately the CFT related arguments applied to models in confined
geometries were suitable only for obtaining at most sub-leading terms
of the asymptotics of the mean cluster size, while the exact formulas
still remained off the scope of this approach. The possibility of
the next step opened in the beginning of 2000s. Then,  a burst of interest
to the $O(1)$ DLM was ignited by an observation by Razumov and Stroganov
of a nice combinatorial structure of the ground state of the XXZ chain
and the six-vertex model at a specific
combinatorial point \cite{RS-01}. A connection of the results of \cite{RS-01} to the $O(1)$ DLM  was pointed at   in 
\cite{BGM}.  A number of relations between
$O(1)$ DLM, the six vertex model, the XXZ model, the fully packed
loop model and alternating sign matrices came from the studies of
this subject \cite{RS-04,RS-05,G}. In particular, several sum rules for the components of the ground state eigenvector  of the $O(1)$ DLM transfer matrix and its generalizations were obtained \cite{FZ2005,FZ2005_2,Z2006,FZZ2006,FZ2007,RSZ2007,CS}. Also statistics of several observables describing connectivity of boundary points
\cite{GBNM,MNGB-1,GierJacobsenPonsaing} and loop embeddings \cite{Mitra Nienhuis} on finite lattices were studied in lattices with different boundary
conditions like cylinder, strip, e.t.c. For many of them exact formulas
were either conjectured or proved.  However, to our knowledge, the
simplest quantities like the mentioned densities of loops aka the
densities of critical percolation clusters are not yet in this list. 

In this Letter we fill this gap. We obtain the exact formulas for
$\nu_{c}(L)$ and $\nu_{nc}(L)$ for any even $L$. To this end we
exploit the connection between the free energies of the $O(n)$ DLM
and the six-vertex model. The latter can be found as the largest eigenvalue
of the corresponding transfer-matrix, of which the derivatives with
respect to the fugacities of contractible and non-contractible loops
yield the mean values of interest. The eigenvalue satisfies the Baxter's
T-Q equation as well as the conjugated T-P equation, which are  the functional
relations between the eigenvalue and two polynomials $Q(x)$ and
$P(x)$ having zeroes on the roots of two systems of Bethe equations.
Both equations were solved for $Q(x)$ and $P(x)$ in the so called
stochastic point, corresponding to the $O(1)$ DLM, by Fridkin, Stroganov,
Zagier (FSZ) in \cite{FSZ-1}. Furthermore, simultaneous differentiation
of the transfer-matrix eigenvalue expressed in terms of $Q(x)$ and
$P(x)$ and of the Wronskian relation between them allows one to exclude
unknown terms and to express the derivatives of the eigenvalue at
the stochastic point via the derivatives of the known $Q(x)$ and $P(x)$ in their arguments \cite{FSZ-2}.
The rest of the work and the main technical challenge of this Letter is
a reduction of the complicated expressions obtained in terms of the
hypergeometric functions evaluated at special values of the parameters
and of the argument to a manageable rational form. It is performed
with the use of Kummer's theorem for hypergeometric function  and its contiguous generalizations.
This is the program we complete in the next sections. Similar calculations
were done also in \cite{S} with the eigenvalue of the XXZ chain
and in \cite{Povolotsky2019} in context of the Raise and Peel model
\cite{GNPR}.

\section{Results }

In this section we present the formulas obtained and compare them
with the results mentioned above. Stating the results, to avoid an
alternate use of the parameters $N$ and $L$, we always use the parameter
$N$ implying that $L=2N$.

\subsection{Contractible loops}

The density of contractible loops is obtained in the form.

\begin{eqnarray}
\nu_{c}(2N) & = & \frac{3\Gamma\left(\frac{N}{2}\right)\Gamma\left(\frac{3N}{2}+\frac{1}{2}\right)}{4\Gamma\left(\frac{3N}{2}\right)\Gamma\left(\frac{N+1}{2}\right)}+\frac{\pi^{2}2^{-2N}3^{2-3N}\Gamma(3N)}{\Gamma\left(\frac{N}{2}+\frac{1}{6}\right)^{2}\Gamma\left(\frac{N}{2}+\frac{5}{6}\right)^{2}\Gamma(N)}-\frac{5}{2}\label{eq:nu_c}\\
 & = & \frac{1}{8},\frac{17}{160},\frac{913}{8960},\frac{3953}{39424},\frac{14569}{146432},\frac{3945737}{39829504},\dots\nonumber 
\end{eqnarray}
In the second line we show the numerical values of the quantity obtained
for $N=1,\dots,6$. One can see that the densities are  rational
numbers. Indeed, using the reflection formula for gamma functions
the l.h.s of (\ref{eq:nu_c}) can be recast in the form of an explicit
rational function of $N$
\[
\nu_{c}(2N)=\frac{2^{-2(N+1)}3^{2-3N}\left(2-(-1)^{N}\right)(3N-1)!}{(N-1)!\left(\frac{5}{6}-\frac{N}{2}\right)_{N}^{2}}+\frac{3\left(\frac{N+1}{2}\right)_{N}}{4\left(\frac{N}{2}\right)_{N}}-\frac{5}{2},
\]
written in terms of the factorials and the Pochhammer symbols $(a)_{n}=a(a+1)\dots(a+n-1)$.
Having no oscillatory factors the formula (\ref{eq:nu_c}) is more
suitable for the asymptotical analysis. Using the Stirling formula
we obtain an asymptotic expansion, which in the first three orders
is 
\begin{equation}
\nu_{c}(2N)=\frac{3\sqrt{3}-5}{2}+\frac{1}{4\sqrt{3}}\left(2N\right)^{-2}-\frac{23}{48\sqrt{3}}(2N)^{-4}+O\left(N^{-6}\right).\label{eq:nu_c asymp}
\end{equation}

\subsection{Non-contractible loops}

The density of non-contractible loops we obtained is

\begin{eqnarray}
\nu_{nc}(2N)&=\frac{2^{2(N-2)}\Gamma(N)}{N\pi^{2}\Gamma(3N)}&\left(3^{3N}\Gamma\left(\frac{N}{2}+\frac{1}{6}\right)^{2}\Gamma\left(\frac{N}{2}+\frac{5}{6}\right)^{2}-\frac{12\pi^{2}\Gamma\left(\frac{3N}{2}\right)^{2}}{\Gamma\left(\frac{N}{2}\right)^{2}}\right)\label{eq:nu_nc}\\
 &  &= \frac{1}{8},\frac{11}{320},\frac{421}{26880},\frac{1403}{157696},\frac{4189}{732160},\frac{952067}{238977024},\dots\nonumber 
\end{eqnarray}
where we again show the first six rational values. The explicitly
rational expression of $\nu_{nc}(2N)$ is
\begin{eqnarray*}
\nu_{nc}(2N)	=\frac{3\times2^{2(N-1)}(N-1)!}{N(3N-1)!}	\left(3^{3N-2}\left(2+(-1)^{N}\right)\left(\frac{5}{6}-\frac{N}{2}\right)_{N}^{2}-\left(\frac{N}{2}\right)_{N}^{2}\right).
\end{eqnarray*}
The Stirling formula applied to (\ref{eq:nu_nc}) to three leading
orders yields 
\begin{equation}
\nu_{nc}(2N)=\frac{1}{\sqrt{3}}(2N)^{-2}-\frac{17}{18\sqrt{3}}\left(2N\right)^{-4}+\frac{1021}{216\sqrt{3}}\left(2N\right)^{-6}+O\left(N^{-8}\right).\label{eq:n_nc}
\end{equation}

\subsection{Comparison to earlier results}

As one could expect, the leading order term in the asymptotic expansion
(\ref{eq:nu_c asymp}) coincides with the asymptotic value of the
critical percolation cluster density on the infinite plane lattice
obtained by Temperley and Lieb \cite{TL} and promoted to the exact
numerical value in \cite{ZiffFinchAdamchik}. The finite size correction
to this value obtained in \cite{KlebanZiff} referred to all (both
contractible and non-contractible) loops. It is to be compared to
the sum of the sub-leading term in (\ref{eq:nu_c asymp}) and the
leading term in (\ref{eq:nu_nc}). Our result differs by the factor
of $2$ from that in \cite{KlebanZiff}. This difference stems from
the distinction of the length scales between the original and the
rotated lattice. Specifically, the result in \cite{KlebanZiff} followed
from the CFT prediction \cite{BloteCardyNightingale1986,Affleck1986}
for the form of universal finite size correction to the specific free
energy on the cylinder of circumference $L$
\[
f_{L}=f_{\infty}-\frac{\pi c}{6}L^{-2},
\]
where, when applied to the lattice, the length is supposed to be measured
in the lattice spacings. As in our case the lattice spacing of the
rotated lattice is $\sqrt{2}$, the length $L$ should be replaced
by $\sqrt{2}N$, which explains the discrepancy. 

The conformal anomaly term used in \cite{AlcarazBrankovPriezzhev RittenbergRogozhnikov}
to evaluate the density of non-contractible loops in $O(n)$ DLM referred
to the original lattice. Indeed, the leading term in (\ref{eq:nu_nc})
exactly coincides with the value that follows from the formula in
\cite{AlcarazBrankovPriezzhev RittenbergRogozhnikov} with parameters
corresponding to unit fugacities of both contractible and non-contractible
loops. The sub-leading term of (\ref{eq:nu_c asymp}) can also be
obtained from \cite{AlcarazBrankovPriezzhev RittenbergRogozhnikov}
by differentiation of the anomaly term with respect to the fugacity
of contractible loops.

\section{From loops to the six-vertex model}

\begin{figure}[h]
\centering{}\includegraphics[width=0.5\textwidth]{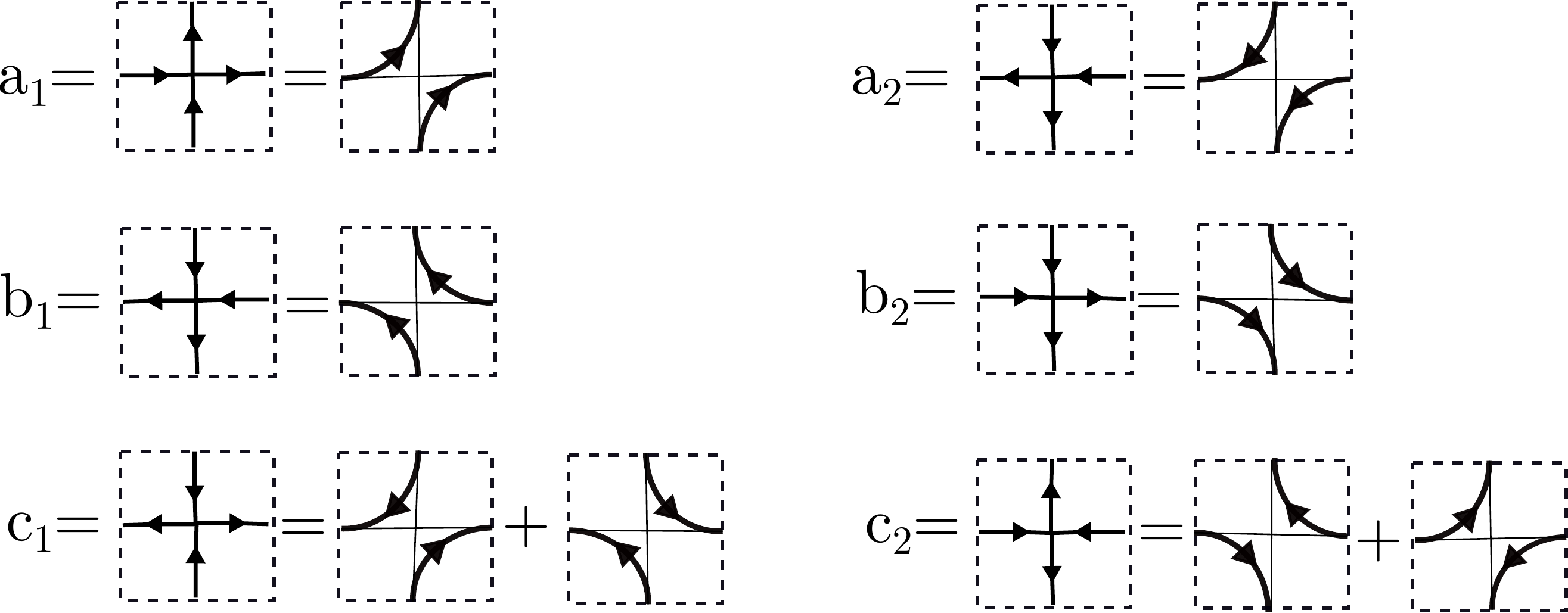}\caption{Correspondence between the six-vertex model and the directed loop
model. \label{fig:six vertex-loop}}
\end{figure}

The solution is based on the relation between the $O(1)$ DLM and
the asymmetric six-vertex model via the directed loop model introduced
as follows \cite{BaxterKellandWu}. We give an orientation to every
loop, which  now can be either clockwise of anti-clockwise. To this
end, we put an arrow in one of two possible directions to each arc
within the vertices in fig. \ref{fig:loop-vertices}. Thus we obtain
eight vertices of the directed loop model. To obtain six vertices
of the six-vertex model out of them we put arrows on the bonds incident
to every site in directions consistent with the directions of the
arcs, as shown in fig. \ref{fig:six vertex-loop}, ignoring the arc
connectivities. Then, two vertices of the six-vertex model will be
the sums of pairs of vertices of the directed loop model and the other
four will be one-to-one. To be short, a summation over the loop orientations
within the directed loop model leads us to the undirected DLM, while
the summation over the arc connectivities, with information about
the arrow directions kept, yields the six-vertex model. 

We assign the following vertex weights using the prescription of \cite{Zinn-Justin2009}.
\begin{eqnarray*}
a_{1} & = & ze^{\mathrm{i}\frac{\varphi}{L}},\quad a_{2}=ze^{-\mathrm{i}\frac{\varphi}{L}},\\
b_{1} & = & e^{-\mathrm{i}\frac{\varphi}{L}},\quad b_{2}=e^{\mathrm{i}\frac{\varphi}{L}},\\
c_{1} & = & zq^{1/2}+q^{-1/2},\quad c_{2}=q^{1/2}+zq^{-1/2}.
\end{eqnarray*}
Here $z$ is an auxiliary spectral parameter that will be set equal
to one in the end. At $z=1$, the weight choice ensures that the contractible
and non-contractible loops come with weights 
\[
w=2_{q}=q+q^{-1}\quad\mathrm{and\quad v=2cos\varphi,}
\]
respectively. In particular we obtain the unit weights, $w=v=1,$
in the so called stochastic point
\begin{equation}
q=e^{\frac{\mathrm{i}\pi}{3}},\quad\varphi=\pi/3,\label{eq:stochastic}
\end{equation}
while the quantities of interest are given by derivatives 
\begin{eqnarray}
\nu_{c} & = & \left.w\frac{d}{dw}\right|_{w=1}f_{L}(w,1),\quad\nu_{nc}=\left.v\frac{d}{dv}\right|_{v=1}f_{L}(1,v)\label{eq:f-deriv}
\end{eqnarray}
of the specific free energy $f_{L}(w,v)$. The latter is equal to
the logarithm of the largest eigenvalue $\Lambda(z)$ of corresponding
row-to-row transfer-matrix normalized to the number of sites $L$
in a horizontal row 
\[
f_{L}(w,v)=L^{-1}\ln\Lambda_{\max}(1).
\]

For the solution of the six vertex model we refer the reader to \cite{Lieb,Sutherland,Baxter}.
To summarize, the transfer matrix describing the transition between
two subsequent horizontal rows of vertical bonds can be written in
a basis of configurations with fixed positions of up-arrows. Since
the action of the transfer matrix preserves the number of up- and
down-arrows, the eigenspace is a direct sum of invariant subspaces
indexed by the number $M$ of up arrows in a horizontal row, which
can take any integer value within the range $0\leq M\leq L$. For
$M$ fixed, we use the Bethe ansatz to diagonalize the transfer matrix
and obtain the eigenvalues in the form 
\begin{equation}
\Lambda(z)=z^{L}e^{\mathrm{i}\varphi}\prod_{i=1}^{M}\frac{1+(z+2_{q})z_{i}e^{-\frac{2\mathrm{i}\varphi}{L}}}{z-z_{i}e^{-\frac{2\mathrm{i}\varphi}{L}}}+e^{-\mathrm{i}\varphi}\prod_{i=1}^{M}\frac{2_{q}z+1+z_{i}ze^{-\frac{2\mathrm{i}\varphi}{L}}}{z_{i}e^{-\frac{2\mathrm{i}\varphi}{L}}-z},\label{eq:eigenvalue}
\end{equation}
 evaluated at numbers $z_{1},\dots,z_{M}$ being the roots of the
Bethe ansatz equations (BAE)
\begin{equation}
z_{j}^{L}=\left(-1\right)^{M-1}\prod_{l=1}^{M}\frac{1+e^{-\frac{2\mathrm{i}\varphi}{L}}2_{q}z_{j}+e^{-\frac{4\mathrm{i}\varphi}{L}}z_{l}z_{j}}{1+e^{-\frac{2\mathrm{i}\varphi}{L}}2_{q}z_{l}+e^{-\frac{4i\varphi}{L}}z_{l}z_{j}},\,\,\,i=1,\dots,M.\label{eq:BAE}
\end{equation}
For further convenience we make a variable change 

\[
z=\frac{u-q}{1-qu},\quad z_{i}=e^{\frac{2\mathrm{i}\varphi}{L}}\frac{u_{i}-q}{1-qu_{i}},
\]
to arrive at the following form of the eigenvalue 
\begin{equation}
\Lambda\left(\frac{u-q}{1-uq}\right)=\left(\frac{u-q}{1-uq}\right)^{L}\frac{e^{\mathrm{i}\varphi}}{\left(-q\right)^{M}}\prod_{i=1}^{M}\frac{uq^{2}-u_{i}}{u-u_{i}}+\frac{\left(-q\right)^{M}}{e^{\mathrm{i}\varphi}}\prod_{i=1}^{M}\frac{uq^{-2}-u_{i}}{u-u_{i}}\label{eq:eigenvalue2}
\end{equation}
and the system of BAE for the parameters $u_{1},\dots,u_{M}$ to be
substituted to (\ref{eq:eigenvalue2})

\begin{equation}
e^{2\mathrm{i}\varphi}\left(\frac{u_{i}-q}{1-qu_{i}}\right)^{L}=\left(-1\right)^{M-1}\prod_{j=1}^{M}\frac{q^{2}u_{j}-u_{i}}{q^{2}u_{i}-u_{j}},\quad i=1,\dots,M.\label{eq:BAE-x}
\end{equation}
Every solution of this system yields a particular eigenvalue. A straightforward
approach would be to find the solution of (\ref{eq:BAE}) corresponding
to the largest eigenvalue, to substitute it into (\ref{eq:eigenvalue}),
and to differentiate. Of course, finding a solution of nonlinear algebraic
system explicitly is not a manageable problem. One, however, can get
around this problem by writing an equation directly for the eigenvalue.
The necessary technique is based on the T-Q equation and its FSZ solution
in the stochastic point.

\section{T-Q equation and FSZ solution}

Alternatively, the eigenvalue problem can be rewritten as a single
functional relation for the polynomial 
\[
Q(u)=\prod_{i=1}^{M}(u-u_{i}),
\]
with the roots being  a particular solution of (\ref{eq:eigenvalue}).
To this end we note that the eigenvalue (\ref{eq:eigenvalue}) must
be polynomial in $z$ of degree at most $L$. Hence the quantity 
\[
T(u)=\Lambda\left(\frac{u-q}{1-uq}\right)(1-uq)^{L}(-q)^{M-L}
\]
 is polynomial in $u$ of at most the same degree. In terms of $T(u)$
the formula (\ref{eq:eigenvalue2}) takes the form

\begin{equation}
T(u)Q(u)=e^{\mathrm{i}\varphi}\phi\left(uq^{-1}\right)Q(uq^{2})+e^{-\mathrm{i}\varphi}\phi\left(uq\right)Q(uq^{-2})(-q)^{2M-L},\label{eq:T-Q-1}
\end{equation}
where we introduced 
\begin{equation}
\phi(u)=\left(1-u\right)^{L}\label{eq:phi}
\end{equation}
and multiplied both sides of (\ref{eq:eigenvalue2}) by the denominator
of its r.h.s. A condition of polynomiality of $T(x)$ suggesting that
the r.h.s. of (\ref{eq:T-Q-1}) is divisible by $Q(u)$, i.e. the
Bethe roots are zeroes of r.h.s. of (\ref{eq:T-Q-1}), is equivalent
to the system (\ref{eq:BAE-x}) of BAE. The idea, however, is to attack
the problem by solving the functional relation (\ref{eq:T-Q-1}) for
two unknown polynomials $T(u)$ and $Q(u)$. 

Before going to the solution we introduce a conjugated problem for
$T(u)$ and another polynomial $P(u)$ of degree $(L-M).$ This problem
arises if we solve the eigenproblem for the transfer-matrix in a different
basis keeping track for positions of the down- instead of the up-arrows.
This is equivalent to solving the original six-vertex model with the
weights $a_{1},b_{1},c_{1}$ and $a_{2},b_{2},c_{2}$ exchanged respectively,
which in practice is realized by the change $\varphi\leftrightarrow-\varphi$.
Since we did nothing but the change of the basis, the solution of
the problem with $(L-M)$ down-arrows should produce the same eigenvalues
$\Lambda(z)$. Thus, for $T(u)$ we obtain 
\begin{equation}
T(u)P(u)=(-q)^{2M-L}e^{\mathrm{-i}\varphi}\phi\left(uq^{-1}\right)P(uq^{2})+e^{\mathrm{i}\varphi}\phi\left(uq\right)P(uq^{-2}).\label{eq:T-P}
\end{equation}
Multiplying eqs. (\ref{eq:T-Q-1}) and (\ref{eq:T-P}) by $P(u)$
and $Q(u)$ respectively, subtracting one from the other and analyzing
the structure of zeroes of the terms of equation obtained, we arrive
at the quantum Wronskian relation between $Q(u)$and $P(u)$.

\begin{equation}
\phi(u)=\frac{e^{\mathrm{i}\varphi}Q(qu)P(q^{-1}u)-e^{-\mathrm{i}\varphi}Q(q^{-1}u)P(qu)(-q)^{2M-L}}{e^{\mathrm{i}\varphi}(-q)^{2M-L}-e^{-\mathrm{i}\varphi}}\label{eq:phi-PQ}
\end{equation}
Substituting this into either of (\ref{eq:T-Q-1}) or (\ref{eq:T-P})
we also obtain
\begin{equation}
T(u)=\frac{e^{\mathrm{2i}\varphi}Q(q^{2}u)P(q^{-2}u)-e^{-2\mathrm{i}\varphi}Q(q^{-2}u)P(q^{2}u)(-q)^{2M-L}}{e^{\mathrm{i}\varphi}(-q)^{2M-L}-e^{-\mathrm{i}\varphi}}.\label{eq:T-PQ}
\end{equation}
These two equations are the key formulas for our solution. 

Now, we are in a position to proceed with the solution for the largest
eigenvalue. This solution corresponds to the choice 
\begin{equation}
	M=N=L/2.
\end{equation}
Thus, the factor $(-q)^{2M-L}$ disappears from the equations (\ref{eq:phi-PQ},\ref{eq:T-PQ}), which then take exactly the form studied in  \cite{Povolotsky2019}. In this case, the solutions of T-Q and T-P equations for the XXZ chain with anisotropy parameter $\Delta=-1/2$ and twisted boundary conditions
corresponding to the stochastic point (\ref{eq:stochastic}) were
found in \cite{FSZ-2}. Here we keep to the notations of \cite{Povolotsky2019}
and refer the reader to the formulas in that paper. The solution for
$T(x)$ has the form 
\begin{equation}
T(x)=(1+x)^{2N}\label{eq: T(x) FSZ}
\end{equation}
that in particular suggests $\Lambda(1)=2^{2N}$, which is simply
the number of possible arrangements of vertices  from fig. \ref{fig:loop-vertices}
within one horizontal row. The polynomials $Q(x)$ and $P(x)$ are
looked for in the form 
\begin{equation}
Q(x)=f_{Q}(x)(1+x)^{-2N},P(x)=f_{P}(x)\left(1+x\right)^{-2N},\label{eq:Q-f_Q,P-f_p}
\end{equation}
where in our case it is convenient to represent the functions $f_{Q}(x)$
and $f_{P}(x)$ in terms of Gauss hypergeometric function (see \cite{Povolotsky2019})

\begin{eqnarray}
f_{Q}(x) & = & \frac{\Gamma(2/3)}{\Gamma(2/3-N)}\left(\frac{\Gamma(2/3)}{\Gamma(2/3+N)}\left._{2}F_{1}(1/3-N,-N,1/3;-x^{3})\right.\right.\label{eq:f_Q}\\
 & + & \left.\frac{x^{2}N\Gamma(-2/3)}{\Gamma(1/3+N)}\left._{2}F_{1}(2/3-N,1-N,5/3;-x^{3})\right.\right),\nonumber 
\end{eqnarray}
\begin{eqnarray}
f_{P}(x) & = & \frac{\Gamma(2/3+N)}{\Gamma(2/3)}\left(\frac{\Gamma(2/3-N)}{\Gamma(2/3)}\left._{2}F_{1}(2/3-N,-N,2/3;-x^{3})\right.\right.\label{eq:f_P}\\
 & + & \left.\frac{xN\Gamma(1/3-N)}{\Gamma(4/3)}\left._{2}F_{1}(1/3-N,1-N,4/3;-x^{3})\right.\right).\nonumber 
\end{eqnarray}

\section{Calculating derivatives.}

The final step is to calculate the derivatives (\ref{eq:f-deriv})
of the free energy with respect to fugacities $w$ and $v$ at $v=w=1$.
This corresponds to differentiation of the eigenvalue with respect
to $q$ and $\varphi$ at the stochastic point. Thus, in terms of
$T(u)$ we have 
\begin{eqnarray}
\nu_{c} & = & \frac{1}{2}+\frac{(1-q^{-2})^{-1}}{2N}\frac{d\ln T(1)}{dq},\quad\nu_{nc}=-\frac{1}{2\sqrt{3}N}\frac{d\ln T(1)}{d\varphi},\label{eq:nu_c-T,nu_nc-T}
\end{eqnarray}
where the coefficients of the derivatives come from the change of variables  $v$ and $w$ in (\ref{eq:f-deriv}) to $q$ and $\varphi$ respectively.

To differentiate $T(1)$ in $q$ we use its expression given by  r.h.s.
of (\ref{eq:T-PQ}). The result 
\[
\frac{d\ln T(1)}{dq}=2^{-2N}\left(2A+B\right),
\]
consists of two parts. One is from the explicit dependence of r.h.s.
of (\ref{eq:T-PQ}) on $q$. The other is from an implicit dependence
of $Q(x)$ and $P(x)$ on $q$ via the $q$-dependence of Bethe roots.
The first one is encapsulated in the letter $A$. It contains the derivatives with respect to $q$   applied to   the arguments of   $Q(q^{\pm 2}u)$ and $P(q^{\pm 2}u)$. The coefficient $2$ of $A$ comes from the exponent of  $q^{\pm  2}$. The quantity  $A$ can explicitly be calculated
 to
\begin{eqnarray}
&&\!\!\!\!\!\!\!\!\!\!\!\!\!\!\!\!\!\!\!\!\!\!\!\!\!\!\!\!A=\frac{qQ'\left(q^{-2}\right)P\left(q^{2}\right)-Q'\left(q^{2}\right)P\left(q^{-2}\right)-q^{-1}Q\left(q^{2}\right)P'\left(q^{-2}\right)-q^{-1}Q\left(q^{-2}\right)P'\left(q^{2}\right)}{q-q^{-1}}\nonumber\\
&&\!\!\!\!\!\!\!\!\!\!\!\!\!\!\!\!\!\!\!\!\!\!\!=\frac{1}{\mathrm{i}\sqrt{3}}\Big\{qf'_Q(q^{-2})f_P(q^2)-f'_Q(q^{2})f_P(q^{-2})-q^{-1}f_Q(q^{2})f'_P(q^{-2}) -q^{-1}f_Q(q^{-2})f'_P(q^2)\Big\}\label{eq:A}\\
&&-2N\Big\{(f_Q(q^{-2})f_P(q^2)-\frac{1+q^{-1}}{\mathrm{i}\sqrt{3}}f_Q(q^2)f_P(q^{-2}))\Big\} \nonumber
\end{eqnarray}
Here and below we assume that $e^{\mathrm{i}\varphi}=q=e^{\mathrm{i}\pi/3}$
and, hence, $q+q^{-1}=1$, $q-q^{-1}=\mathrm{i}\sqrt{3}$ and $q^{3}=q^{-3}=-1,q^{2}=q^{-4}=-q^{-1}=-q^{5}$, e.t.c. In particular, after differentiating we replace 
$e^{\pm \mathrm{i}\varphi}$ by $q^{\pm 1}$ respectively.
The second part, expressed via the letter $B$, is unknown. However,
$B$ can be found from yet another derivative of eq. (\ref{eq:phi-PQ})
\[
\left.\frac{d\phi(u)}{dq}\right|_{u=q^{3}}=-A+B=0,
\]
which apparently vanishes on one hand and includes $A$ and $B$ on
the other. Note that we first differentiate and then substitute $u=q^{3}=-1$.
As a result we obtain 

\begin{eqnarray}
\frac{d\ln T(1)}{dq} & = & 3\times2^{-2N}A.\label{eq:dlnT/dq}
\end{eqnarray}
Similarly for the derivatives in $\varphi$ we have 
\[
\frac{d\ln T(1)}{d\varphi}=\frac{1}{\sqrt{3}}\left(-1+\frac{2C+D}{2^{2N}}\right),\quad\left.\frac{d\phi(u)}{d\varphi}\right|_{u=q^{3}}=\frac{1}{\sqrt{3}}\left(D-C-2^{2N}\right)=0,
\]
with 
\begin{eqnarray}
C&=&q^{2}Q\left(q^{2}\right)P\left(q^{-2}\right)+q^{-2}Q\left(q^{-2}\right)P\left(q^{2}\right)\label{eq: C}\\
&=&q^2f_Q(q^2)f_P(q^{-2})+q^{-2}f_Q(q^{-2})f_P(q^{2}),\nonumber
\end{eqnarray}
coming from the explicit dependence on $\varphi$ of the numerators of $T(u)$ and $\phi(u)$ and $D$ being the unknown part. There are also  constants not included neither in $C$ nor in $D$, which came from  differentiating  the denominators. 
Then we obtain 
\begin{equation}
\frac{d\ln T(1)}{d\varphi}=\frac{\sqrt{3}C}{2^{2N}},\label{eq:dlnT/dphi}
\end{equation}
which agrees with formula obtained previously in \cite{FSZ-2} in
context of XXZ chain with imaginary magnetic field.

In fact, eqs. (\ref{eq:nu_c-T,nu_nc-T}-\ref{eq:dlnT/dphi}) together
with formulas (\ref{eq:Q-f_Q,P-f_p}-\ref{eq:f_P}) already provide
explicit answers. However, they are not yet of satisfactory form being
sums of hardly computable  terms. For example, consider the  four quantities   $f_Q(q^{\pm2}),f_P(q^{\pm2})$ to be substituted to ({\ref{eq:A},\ref{eq: C}}). Obtained from (\ref{eq:f_Q},\ref{eq:f_P}) by setting  $x=q^{\pm 2}$ each of them consists of two terms containing the hypergeometric function 
$_2F_1(a,b,c;t)$ with  some  parameters $a,b,c$, evaluated at $t=-x^3=-1$.     
%\begin{eqnarray}
%f_{Q}(q^{\pm 2}) & = & \frac{\Gamma(2/3)}{\Gamma(2/3-N)}\left(\frac{\Gamma(2/3)}{\Gamma(2/3+N)}\left._{2}F_{1}(1/3-N,-N,1/3;-1)\right.\right.\label{eq:f_Q_1}\\
%& - & \left.\frac{q^{\mp 2}N\Gamma(-2/3)}{\Gamma(1/3+N)}\left._{2}F_{1}(2/3-N,1-N,5/3;-1)\right.\right),\nonumber \\
%f_{P}(q^{\pm 2}) & = & \frac{\Gamma(2/3+N)}{\Gamma(2/3)}\left(\frac{\Gamma(2/3-N)}{\Gamma(2/3)}\left._{2}F_{1}(2/3-N,-N,2/3;-1)\right.\right.\label{eq:f_P_1}\\
%& + & \left.\frac{q^{\pm 2} N\Gamma(1/3-N)}{\Gamma(4/3)}\left._{2}F_{1}(1/3-N,1-N,4/3;-1)\right.\right).\nonumber 
%\end{eqnarray}
Likewise, by  direct differentiation of (\ref{eq:f_Q},\ref{eq:f_P}) with the use of formula
\begin{equation}
\partial\left._{2}F_{1}\right.(a,b,c;t)/\partial t=abc^{-1}\,_{2}F_{1}(a+1,b+1,c+1;t)\label{eq: derivative}
\end{equation} 
the derivatives $f'_Q(q^{\pm2})$ and $f'_P(q^{\pm 2})$ 
necessary for (\ref{eq:A}) can be calculated to  similar three term expressions. 
The formulas obtained in this way are not   very informative. In particular, they are not  suitable for the asymptotic analysis. 
Our next goal is  to transform them to a more tractable form. 

To this end, we need to evaluate the hypergeometric functions $_2F_1(a,b,c;-1)$ that appear in $f_Q(q^{\pm2}),f_P(q^{\pm 2})$ and in $f'_Q(q^{\pm2}),f'_P(q^{\pm 2})$. The well known Kummer's theorem \cite{AndrewsAskeyRoy} gives such an evaluation
resulting in a ratio of gamma functions, provided that the  relation $c=1+a-b$ between the parameters holds. One can see that the parameters of hypergeometric
functions in (\ref{eq:f_Q},\ref{eq:f_P}) satisfy    relations $1+a-b-c=\pm1$ shifted by $\pm1$. In addition, according to (\ref{eq: derivative}), the derivatives $f'_Q(q^{\pm2}),f'_P(q^{\pm 2})$  in (\ref{eq:A})  will contain the   hypergeometric functions  with the argument $t=-1$ and parameters satisfying  the original Kummer's relation as well as  the relations  $1+a-b-c=\pm 2$   shifted by $\pm 2$. Luckily,  contiguous
generalizations of the Kummer's theorem proved in \cite{ChoiRathieMalani}
are applicable to these cases. They express the hypergeometric functions at $t=-1$
with parameters satisfying  shifted  relations as  sums of  ratios of
gamma functions as follows 
\begin{equation}
\!\!\!\!\!\!\!\!\!\!\!\!\!\!\!\!\!\!\!\!\!\!\!\!\!\!\!\!\!\!\!\!\!\!{}_{2}F_{1}(a,b,1+a-b+n;-1)=\frac{\Gamma(1+a-b+n)\Gamma(1-b)}{2\Gamma(a)\Gamma(1-b+n)}\sum_{k=0}^{n}(-1)^{k}\left(\begin{array}{c}
n\\
k
\end{array}\right)\frac{\Gamma\left(\frac{a}{2}+\frac{k}{2}\right)}{\Gamma\left(\frac{a}{2}+\frac{k}{2}-b+1\right)},\label{eq: Kummer's cont_1}
\end{equation}
\begin{equation}
\!\!\!\!\!\!\!\!\!\!\!\!\!\!\!\!\!\!\!\!\!\!\!\!\!\!\!\!\!\!\!\!\!\!{}_{2}F_{1}(a,b,1+a-b-n;-1)=\frac{\Gamma(1+a-b-n)\Gamma(1-b-n)}{2\Gamma(a)\Gamma(1-b-n)}\sum_{k=0}^{n}\left(\begin{array}{c}
n\\
k
\end{array}\right)\frac{\Gamma\left(\frac{a}{2}+\frac{k}{2}\right)}{\Gamma\left(\frac{a}{2}+\frac{k}{2}-b+1-n\right)},\label{eq: Kummer's cont_2}
\end{equation}
where $n\in\mathbb{N}_{0}$. For example, applying this to (\ref{eq:f_Q},\ref{eq:f_P}) we obtain
\begin{eqnarray}
\!\!\!\!\!\!\!\!\!\!\!\!\!\!\!\!\!\!\!\!\!\!\!\!\!f_{Q}(q^{\pm2})=	\frac{\pi}{\sqrt{3}}\frac{\Gamma\left(\frac{2}{3}\right)}{\Gamma\left(\frac{2}{3}-N\right)}\Biggl[\frac{1}{\Gamma\left(\frac{2}{3}+N\right)\Gamma\left(\frac{1}{3}-N\right)}\left(\frac{\Gamma\left(\frac{1}{6}-\frac{N}{2}\right)}{\Gamma\left(\frac{1}{6}+\frac{N}{2}\right)}+\frac{\Gamma\left(\frac{2}{3}-\frac{N}{2}\right)}{\Gamma\left(\frac{2}{3}+\frac{N}{2}\right)}\right)\nonumber\\
\,\,\,\,\,\,\,\,\,\,\,\,\,\,\,\,\,\,\,\,\,\,\,\,\,\,\,\,\,\,-\frac{q^{\mp2}}{\Gamma\left(\frac{1}{3}+N\right)\Gamma\left(\frac{2}{3}-N\right)}\left(\frac{\Gamma\left(\frac{1}{3}-\frac{N}{2}\right)}{\Gamma\left(\frac{1}{3}+\frac{N}{2}\right)}+\frac{\Gamma\left(\frac{5}{6}-\frac{N}{2}\right)}{\Gamma\left(\frac{5}{6}+\frac{N}{2}\right)}\right)\Biggr]\nonumber\\
\!\!\!\!\!\!\!\!\!\!\!\!\!\!\!\!\!\!\!\!\!\!\!\!f_{P}(q^{\pm2})=\frac{\Gamma\left(\frac{2}{3}+N\right)}{2\Gamma\left(\frac{2}{3}\right)}\Biggl[\left(\frac{\Gamma\left(\frac{1}{3}-\frac{N}{2}\right)}{\Gamma\left(\frac{1}{3}+\frac{N}{2}\right)}+\frac{\Gamma\left(\frac{5}{6}-\frac{N}{2}\right)}{\Gamma\left(\frac{5}{6}+\frac{N}{2}\right)}\right)+q^{\pm2}\left(\frac{\Gamma\left(\frac{1}{6}-\frac{N}{2}\right)}{\Gamma\left(\frac{1}{6}+\frac{N}{2}\right)}+\frac{\Gamma\left(\frac{2}{3}-\frac{N}{2}\right)}{\Gamma\left(\frac{2}{3}+\frac{N}{2}\right)}\right)\Biggr]\nonumber
\end{eqnarray}
where we also used the reflection identity 
\begin{equation}
\Gamma(z)\Gamma(1-z)=\frac{\pi}{\sin\pi z} \label{eq: reflection}
\end{equation} to reduce the number of  gamma functions.
Similar though longer formulas can also be  obtained  for $f'_{Q}(q^{\pm2})$  and $f'_{P}(q^{\pm2})$. Substituting the eight obtained expressions   into  (\ref{eq:A},\ref{eq: C})  we arrive at large  sums of rational combinations of  gamma  functions. The remaining
part of the work though tedious is straightforward. It consists in step by step simplification of the obtained expressions with the  use of the  reflection identity (\ref{eq: reflection}), the Legendre duplication formula
\begin{equation}
	\Gamma(z)\Gamma(z+1/2)=\sqrt{\pi}2^{1-2z}\Gamma(2z)
\end{equation}
and the  trigonometric identities. At every step we try to  reduce the number of gamma functions and trigonometric functions appearing from the reflection identity.  Finally, we arrive at the shortest form we are able to obtain, which being   substituted to (\ref{eq:dlnT/dq},\ref{eq:dlnT/dphi}) and then to (\ref{eq:nu_c-T,nu_nc-T}) gives us   formulas (\ref{eq:nu_c},\ref{eq:nu_nc}).

\section{Discussion and outlook}

To summarize we have obtained the densities of contractible and non-contractible
loops on the square lattice and of critical percolation clusters
on the forty-five degree rotated square lattice, both rolled into
a cylinder. To this end, we used the mapping of the dense loop model
to the six-vertex model, and the method of solution of the T-Q and
T-P equations proposed earlier by Fridkin, Stroganov and Zagier. The
results must also be related to some correlation functions over the
stationary state of the associated Markov chain like it was e.g. in
context of its continuous-time analogue Raise and Peel model \cite{Povolotsky2019}.
Which one, however, is not obvious and requires careful investigation
in the more complex discrete framework of the $O(1)$ dense loop model
.

It has been shown that the leading orders of the asymptotic expansion
reproduce previous results. In particular, the sub-leading order of
the density of contractible loops and the leading order of the density
of non-contractible loops are known to be universal having a meaning
within the conformal field theory. The exact formulas obtained here
allow in principle deriving the asymptotic expansion to any finite
order. It would be interesting to understand whether the higher order
finite size corrections can be combined into quantities with any degree
of universality \cite{Privman1991}. 

The used technique must be applicable to a wider class of models.
In particular the T-Q equation for the XXZ model with free boundary
conditions related to the dense loop model on the strip was solved
in \cite{FSZ-2} and a quantity similar to our (\ref{eq:dlnT/dq})
was found in terms of $Q$- and $P$-polynomials. No final explicit
formulas like those presented here were however obtained. Another
direct generalization is to consider the lattice with odd $L$. In
this case  one infinite path on the cylinder exists
and no non-contractible loops present. The technique used here was
extended to odd $L$ in \cite{S}, where it was applied to the eigenvalue
of the Hamiltonian of the XXZ model. No results for the discrete lattice
model were yet obtained. 

More ambitious problems would be an extension of the technique to
obtain exact higher cumulants of loop densities, which were also asymptotically
predicted from the conformal field theory arguments in \cite{KlebanZiff}.
It would also be of interest to find extensions to loop models related
with higher spin \cite{Nienhuis1990} or rank \cite{FendleyJacobsen2008}
integrable systems. Another challenge is the  generalization  of the technique used here
to integrable models at the elliptic level like e.g. eight-vertex model and XYZ spin chain. In principle  the parameters  analogous to our $q$ and $\varphi$ of the six vertex model  can also be introduced in that case \cite{BM1} as well as the technique of  T-Q equation is  applicable. This makes potentially possible to evaluate the conjugate ground state observables  as the derivatives of the largest eigenvalue with respect to these parameters. Also, there is an analogue of the combinatorial point \cite{RS2010} that shows similarly nice properties and  admits obtaining  exact analytic results about the ground state   eigenvalue \cite{BM2} and eigenvector \cite{BM3,Z2013}. 
Still it is not known whether an analogue of the FSZ method exists in that case.  Also there is no  interpretation of the eight-vertex model in terms of loops. Is there an appropriate generalization of the loop picture for that case is an interesting question. The study of these issues in both analytic and combinatorial perspectives is the matter for further investigation.

\ack
The work is supported by Russian Foundation of Basic Research under
grant 20-51-12005.

\end{document}